\documentclass[12pt]{article}
\usepackage{cite}
\usepackage{graphicx}
\usepackage{amssymb,amsmath,amsfonts,amsthm}
\usepackage{epstopdf}
\usepackage[normalem]{ulem}
\usepackage{comment}
\usepackage{hyperref}

\hypersetup{
   colorlinks=true,         
   linkcolor=blue,          
   citecolor=red,           
   urlcolor=Violet          
}

\usepackage[usenames,dvipsnames]{color}

\begin{document}

\title{\Large \bf{Matching lightcone- and anomaly-sum-rule predictions for the
                  pion-photon transition form factor} }

\author{
A.~G.~Oganesian,$^{1,2}$
A.~V.~Pimikov,$^{1,3}$ N.~G.~Stefanis,$^4$
O.~V.~Teryaev,$^1$
\footnote{
  Electronic addresses:
  \href{mailto:armen@itep.ru,}{armen@itep.ru},
  \href{mailto:pimikov@theor.jinr.ru}{pimikov@theor.jinr.ru},
  \href{mailto:stefanis@tp2.ruhr-uni-bochum.de}{stefanis@tp2.ruhr-uni-bochum.de},
  \href{mailto:teryaev@theor.jinr.ru}{teryaev@theor.jinr.ru}.
}
 \vspace{12pt} \\
\it \small $^1$ Bogoliubov Laboratory of Theoretical Physics,\\
\it \small Joint Institute for Nuclear Research, 141980, Dubna, Russia,\\
\it \small $^2$Institute of Theoretical and Experimental Physics, 117218,  Moscow, Russia\\
\it \small $^3$Institute of Modern Physics, Chinese Academy of Sciences,
               Lanzhou, 730000, China\\
\it \small $^4$Institut f\"{u}r Theoretische Physik II,
           Ruhr-Universit\"{a}t Bochum,
           D-44780 Bochum, Germany
       }

\date{}

\hfill RUB-TPII-03/2015
{\let\newpage\relax\maketitle}


\begin{abstract}
The pion-photon transition form factor is studied by employing two types
of Sum Rules: Light Cone Sum Rules (LCSR) and Anomaly Sum Rules (ASR).
By comparing the predictions for the pion-photon transition form factor,
obtained from these two approaches, the applicability limit of the LCSRs
at low momenta is determined.
Reciprocally, the ASR threshold dependence on the momentum was extracted
using our LCSR-based method in combination with two different types of
pion distribution amplitudes and found that at higher $Q^2$ it approaches
a constant.

\end{abstract}

\newpage

\section{Introduction}
\label{sec:intro}
The measurement of the transition form factor (TFF)
$ F_{\pi\gamma}(Q^{2})$ at  $4<Q^2<40$~GeV$^2$
by the BaBar Collaboration \cite{Aubert:2009mc} in 2009 has showed very
unexpected results:
although at $Q^2<10$~GeV$^2$ the collected data are in agreement with
previous experiments, the trend of the measured TFF at
$Q^2\gtrsim 10$~GeV$^2$
strongly exceeds the predicted asymptotic limit
\cite{Lepage:1980fj,Brodsky:1981rp}.
This is given by
\begin{equation}
  F_{\pi\gamma}^{\mathrm{asy}}(Q^{2})
=
  \frac{\sqrt{2} f_{\pi}}{Q^{2}}+\mathcal{O}(1/Q^4) \, ,
\label{eq:AsLargeQ}
\end{equation}
and deviations from it are challenging the validity of the factorization
property of hard exclusive processes within Quantum Chromodynamics (QCD).
On the other hand, the more recent data of the Belle Collaboration
\cite{Uehara:2012ag} of the year 2012 do not indicate such a large growth
of the scaled TFF at high $Q^2$.
This significant difference in the data trend, stimulated a number of
theoretical investigations of the TFF at various momentum transfers $Q^2$
some questioning the validity of the BaBar data, e.g., \cite{MS09},
while other proposals attempting to rationalize this peculiar TFF
behavior, for example, \cite{Radyushkin:2009zg,Polyakov:2009je}.

Perturbative Quantum Chromodynamics (pQCD) in the leading-order (LO)
approximation of the collinear factorization approach to the pion TFF
predicts \cite{Lepage:1980fj,Brodsky:1981rp}:
$
 Q^{2}F_{\pi\gamma}(Q^{2})
=
 \left(\sqrt{2}f_\pi/3\right)
 \int_{0}^{1}dx
\varphi_{\pi}^{(2)}\left(x,Q^2\right)/x
=
 \left(\sqrt{2}f_\pi/3\right) \langle x^{-1} \rangle_{\pi}
$,
where $\varphi_{\pi}^{(2)}(x,\mu^2)$
is the pion distribution amplitude (DA) of twist two encoding the
nonperturbative partonic interactions.
This important relation imposes a crucial constraint on the profile
of the pion DA in terms of its inverse moment and holds in any
theoretical framework based on collinear QCD factorization ---
see \cite{Stefanis:2015qha} for a comprehensive discussion.
However, it is not known at which momentum value, ``asymptotia''
is effectively reached and the TFF starts to scale with $Q^2$.

Expressing the DA in terms of the conformal Gegenbauer expansion,
\begin{equation}
  \varphi_{\pi}^{(2)}(x,\mu^2)
=\sum_{n=0,2,4, \ldots}^{\infty} a_{n}(\mu^2) \psi_n(x) \, ,
\label{eq:gegen-exp}
\end{equation}
where $\psi_n(x)=6x(1-x) C^{(3/2)}_n(2x-1)$
with $\varphi_{\pi}^{\rm asy}(x)=6x(1-x)\equiv 6x\bar{x}$
are the eigenfunctions of the
Efremov-Radyushkin-Brodsky-Lepage (ERBL) evolution equation
\cite{ER80,Lepage:1980fj}, one can evolve the TFF at any higher scale
$Q^2\geq\mu^2$ by determining the nonperturbative coefficients
$a_{n}(\mu^2)$ at the scale of choice.
At infinitely large $Q^2$, the pion DA assumes its asymptotic
form $\varphi_{\pi}^{\text{asy}}\left(x\right)$ giving rise to the
asymptotic result for the TFF, viz., Eq.\ (\ref{eq:AsLargeQ}),
in which all possible perturbative and nonperturbative corrections
are absent.
To account for these corrections at finite $Q^2$ values, one has to
apply more sophisticated approaches, like lightcone sum rules (LCSR),
developed in\cite{BBK89,Kho99} and applied by
Bakulev, Mikhailov, Pimikov, and Stefanis (BMPS) in
\cite{BMS02,Bakulev:2003cs,BMS05lat,Mikhailov:2010ud,Bakulev:2011rp,BMPS12,%
      Stefanis:2012yw}
(see also \cite{Khodjamirian:2009ib,ABOP10,ABOP12}).

In an independent parallel development by Klopot, Oganesian, and
Teryaev (KOT)
\cite{Klopot:2010ke,Klopot:2011qq,Klopot:2011ai,Klopot:2012hd},
it was shown that the photon transition form factors of pseudoscalar
mesons can be studied by means of anomaly sum rules (ASRs) which are
based on the dispersive representation of the axial anomaly.
This procedure is closely connected to the treatment of the
vector-vector-axial triangle graph amplitude
\cite{Dolgov:1971ri,Horejsi:1985qu,Horejsi:1994aj},
in which the axial current is assumed to represent the pion and
the two vector currents represent the real and the virtual photon.
The key element of the ASR method is that it does not rely upon the
factorization hypothesis and, in this sense, it is directly related
to the first principles of QCD and the Dynamical Chiral Symmetry
Breaking (DCSB) and the concomitant mass generation of hadrons.
In addition, KOT have extended the analytic continuation of the
ASRs (and herewith the TFF computation) to the timelike region
\cite{Klopot:2013laa}.
This is particularly important because the low-momentum timelike regime
is unreachable in the conventional pQCD approach.

In this paper we want to compare the two approaches, i.e., ASRs vs.
LCSRs, and match their predictions for the pion-photon TFF.
The scope of the analysis is to identify the treacherous points of each
of these methods and determine their accuracy limits.
The paper is organized as follows.
In the next two sections, we will briefly recall the basic ingredients
of both methods, starting with the LCSRs in Sec.\ \ref{sec:lcsr} and
continuing with the ASRs in Sec.\ \ref{sec:asr}.
The comparison of the predictions obtained with the two methods will be
addressed in Sec.\ \ref{sec:lcsr-asr}, while Sec.\ \ref{sec:concl} is
devoted to our conclusions.

\section{LCSR approach}
\label{sec:lcsr}
The behavior of the TFF can be obtained from a detailed formalism
(called in the following BMPS for short),
developed in a series of papers
\cite{BMS01,BMS02,Bakulev:2003cs,BMS05lat,Mikhailov:2010ud,Bakulev:2011rp,BMPS12,Stefanis:2012yw}.
This formalism combines the dispersive approach of LCSRs pioneered in
\cite{BBK89,Kho99} (see also \cite{SY99,ABOP10,ABOP12})
with QCD sum rules which employ nonlocal condensates (NLC)s
\cite{MR86,MR86ev,MR89,MR90,MS93,BR91,MR92}.
The NLC QCD SRs are used to derive the pion DA, while the LCSRs serve
our purpose twofold: first, to take into account the hadronic content
of the low-virtuality photon and second, to incorporate
contributions from QCD perturbation theory and higher-twist corrections.
Within QCD, the pion-photon transition form factor for two off-shell
photons
$F^{\gamma^{*}\gamma^{*}\pi^0}$ is given by the matrix element
\begin{equation}
  \int d^{4}z e^{-iq_{1}\cdot z}
  \langle
         \pi^0 (P)| T (J_{\mu}(z) J_{\nu}(0))| 0
  \rangle
=
  i\epsilon_{\mu\nu\alpha\beta}
  q_{1}^{\alpha} q_{2}^{\beta}
  F^{\gamma^{*}\gamma^{*}\pi^0}(Q^2,q^2)\ ,
\label{eq:matrix-element}
\end{equation}
where $J_\mu$ is the quark electromagnetic current and both photons
are assumed to have finite virtualities
$q_{1}^{2}=-Q^{2}>>q_{2}^{2}=q^{2}>0$.
The LCSR \cite{BBK89,Kho99} for this matrix element reads
\begin{eqnarray}
  Q^2 F^{\gamma^*\gamma^*\pi}\left(Q^2,q^2\right)
& = &
  \frac{\sqrt{2}}{3}f_\pi
  \left[
        \frac{Q^2}{m_{\rho}^2+q^2}
        \int_{x_{0}}^{1}
        \exp\left(
                  \frac{m_{\rho}^2-Q^2\bar{x}/x}{M^2}
            \right)
        \bar{\rho}(Q^2,x)
  \frac{dx}{x} \right.
\nonumber \\
&&
  + \left.
  \rule{0in}{0.25in}
      \int_{0}^{x_0} \bar{\rho}(Q^2,x)
        \frac{Q^2dx}{\bar{x}Q^2+x q^2}
  \right]
\, .
\label{eq:LCSR-FQq}
\end{eqnarray}
Here, the integration limits are defined by
$x_0=Q^2/\left(Q^2+s_{\rho}\right)$ and $s =\bar{x}Q^2/x$,
where $s_{\rho}\simeq 1.5~\text{GeV}^2$ is the effective threshold
in the vector channel, and $M^2$ is the Borel parameter, with
$m_\rho=0.77$~GeV denoting the physical mass of the $\rho$ meson.
The main theoretical ingredient in the above sum rule is
the spectral density
$
 \bar{\rho}(Q^2,x)=(Q^2+s)\rho^\text{pert}(Q^2,s)
$,
where
\begin{eqnarray}
  \rho^\text{pert}(Q^2,s)
=
  \frac{1}{\pi} {\rm Im}F^{\gamma^*\gamma^*\pi^0}
  \left(Q^2,-s-i\varepsilon\right)
=
  \rho_\text{tw-2}
  +\rho_\text{tw-4}
  +\rho_\text{tw-6}
  +\ldots \, .
\label{eq:rho-twists}
\end{eqnarray}

The leading-twist spectral density was studied up to the
next-to-next-to-leading order (NNLO) level of pQCD, i.e.,
$
 \rho_\text{tw-2}
=
 \rho_\text{LO}+\rho_\text{NLO}+\rho_{\text{NNLO}_{\beta_0}}+\cdots
$,
albeit only the $\beta_0$ part of the NNLO term
is known \cite{Melic:2002ij} (see also \cite{MS09}).
For the next-to-leading-order (NLO) term $\rho_{\rm NLO}$, we employ
the expression computed in \cite{MS09} with the correction pointed out
in \cite{ABOP10}, whereas the leading-order contribution is given by
the Born term.
To compute the spectral density, one takes recourse to the
hard-scattering amplitudes, which are calculable within pQCD as a
series expansion in terms of the coupling parameter
$a_{s}(\mu_{\rm R}^{2})=\alpha_{s}(\mu_{\rm R}^{2})/(4\pi)$,
where $\mu_{\rm R}^{2}$ is the renormalization scale.
In the works of the BMPS team, the renormalization and the
factorization scale have been identified and set equal to $Q^2$.
This choice avoids the appearance of scheme-dependent numerical
coefficients which play no important role in our considerations.
All said, the leading twist-two expression for the pion-photon TFF
has the perturbative expansion
\begin{eqnarray}
\label{eq:T_exp}
  F_{\gamma^*\gamma^*\pi^0}^\text{tw-2}
\sim
 \left[
        T_\text{LO}
       +
        a_s(\mu^2) T_\text{NLO}
       +
        a_s^2(\mu^2) T_{\text{NNLO}_{\beta_0}}
       +
        \ldots
 \right]
\otimes
       \varphi_{\pi}^{(2)}(x,\mu^2) \, ,
\end{eqnarray}
where $\otimes \equiv \int_{0}^{^1}dx$.
The TFF is dominated by the first nontrivial perturbative term
proportional to $T_\text{NLO}$ and the twist-four contribution.
On the other hand, the overall uncertainties have various sources.
These are:
(i) uncertainties related to the particular pion DA model adopted,
(ii) twist-four uncertainties,
(iii) estimated uncertainties related to the twist-two contribution
at the $\text{NNLO}_{\beta_0}$ level, and
(iv) uncertainties induced by the twist-six term.
The latter two uncertainties are taken into account by means of
their sum because for $M^2\approx 0.75$~GeV$^2$
they are comparable in size and have a relatively small magnitude
but enter with opposite signs \cite{Bakulev:2011rp}.
For larger values of the Borel parameter around $M=1.5$~GeV$^2$, as
used for instance in the LCSR analysis in \cite{ABOP10}, the
uncertainties related to the twist-six term turn out to be small.

For definiteness, we use for the calculation of the TFF the family of
the pion DAs determined in \cite{BMS01} with the help of QCD sum rules
with NLCs and the nonlocality parameter $\lambda_q^2=0.4$~GeV$^2$.
In addition, we also employ the shorttailed platykurtic pion DA, which
was more recently derived and discussed in
\cite{Stefanis:2014nla,Stefanis:2014yha,Stefanis:2015qha} using the same
NLC QCD SR method but with a slightly larger value of the
nonlocality parameter, notably, $\lambda_q^2=0.45$~GeV$^2$.
The platykurtic model DA is unimodal, like the asymptotic DA, but is
much broader than this and has its tails ($x=0,1$) suppressed like
the classic bimodal BMS DA.
It combines intrinsically the characteristic features of the
$x$-distribution of the valence quark in the pion bound state being
subject to DCSB --- as described via Dyson-Schwinger (DS) equations
\cite{Chang:2013pq,Gao:2014bca} --- with correlations induced
by NLCs in a nontrivial vacuum which give rise to a finite vacuum
quark virtuality.
The first effect, related to the mass dressing of the confined quark
propagator, causes a broad (unimodal) downward concave shape of the
DA with enhanced tails, while the second effect tends to suppress the
endpoint regions and create two separated peaks.
The reasons why the endpoint-regions of the pion DA should be suppressed,
has been discussed in detail long ago, see e.g., \cite{SSK99,SSK00},
and more recently in \cite{Stefanis:2014nla,Stefanis:2015qha}
having recourse to the synchronization properties of nonlinear
oscillators.
In our estimates, shown in graphical form farther below, the BMS model
DA is used to obtain the central TFF prediction, while the validity
range of the BMS DAs \cite{BMS01} serves as a measure to estimate the
involved intrinsic errors.
Note that the ``platykurtic'' prediction for the pion-photon TFF almost
coincides with that derived with the BMS model DA
\cite{Stefanis:2014nla,Stefanis:2014yha}.

\section{ASR approach}
\label{sec:asr}
In this section we briefly recall the main points of the ASR approach.
This method relies upon the analysis of the vector-vector-axial current
triangle graph amplitude
\begin{equation}
  T_{\alpha \mu\nu}(k,q)
=
  \int d^4 x d^4 y e^{i(k\cdot x+q\cdot y)}
  \langle 0|T\{J_{\alpha 5}^{3}(0) J_\mu (x)J_\nu(y) \}|0\rangle \, ,
\label{eq:VVA}
\end{equation}
where $k$ and $q$ are the momenta of the two photons,
$J_{\alpha 5}^{3}$ is an isovector axial current,
and $J_\mu$ and $J_\nu$ denote the electromagnetic currents.
In what follows, we limit ourselves to the case when
one of the photons is on-shell ($k^2=0$).

It is convenient to write the tensor decomposition of this correlator
in the form (see \cite{Horejsi:1994aj} for details)
\begin{eqnarray}
\label{eq1} \nonumber T_{\alpha \mu \nu} (k,q) & = & F_{1} \;
\varepsilon_{\alpha \mu \nu \rho} k^{\rho} + F_{2} \;
\varepsilon_{\alpha \mu \nu \rho} q^{\rho}
\\ \nonumber
& & + \; \; F_{3} \; k_{\nu} \varepsilon_{\alpha \mu \rho \sigma}
k^{\rho} q^{\sigma} + F_{4} \; q_{\nu} \varepsilon_{\alpha \mu
\rho \sigma} k^{\rho}
q^{\sigma}\\
& & + \; \; F_{5} \; k_{\mu} \varepsilon_{\alpha \nu
\rho \sigma} k^{\rho} q^{\sigma} + F_{6} \; q_{\mu}
\varepsilon_{\alpha \nu \rho \sigma} k^{\rho} q^{\sigma},
\end{eqnarray}
where the coefficients
$F_{j} = F_{j}(k^{2}, q^{2}, p^{2}; m^{2})$,
$p = k+q$, $j = 1, \dots ,6$
are the corresponding Lorentz invariant amplitudes, constrained by
current conservation and Bose symmetry.
Note that the latter includes the interchange
$\mu \leftrightarrow \nu, k \leftrightarrow q$
in the tensor structures and $k^2 \leftrightarrow q^2$ in the
arguments of the scalar functions $F_{j}$.

It was shown in \cite{Horejsi:1994aj} that the invariant amplitude
$F_3$ satisfies the unsubtracted dispersion relation pertaining to
the substraction of the axial-current divergence.
(This seemingly controversial situation is due to the extra factor
$q^2$ in the current divergence.)
Using this unsubtracted dispersion relation, the dispersion
representation of the axial anomaly for any virtual photon amounts to
the ASR \cite{Horejsi:1994aj}
\begin{equation}
\int_{0}^{\infty} A_{3}(s,Q^2; m_i^2) ds =
\frac{1}{2\pi}N_c C \, ,
\label{eq:asr}
\end{equation}
where $N_c=3$ is the number of colors,
$C=\frac{1}{3\sqrt{2}}$ is the charge factor,
and $m_i$ are the quark masses.
Here $Q^2=-q^2$ denotes the momentum transfer of the virtual photon,
whereas $A_3$ is the imaginary part of the invariant amplitude $F_3$.

Although the spectral density $A_{3}(s,Q^{2};m^{2})$
can in principle comprise both, perturbative as well as nonperturbative
corrections, Eq.\ (\ref{eq:asr}) is an exact expression
and receives on its right-hand side neither perturbative corrections,
by virtue of the Adler-Bardeen theorem \cite{Adler:1969gk},
nor nonperturbative contributions, the latter due to
't~Hooft's principle \cite{'tHooft:1980xb}.\footnote{It is worth
mentioning that the first-order correction $\propto \alpha_s$ 
to the integrand itself is zero --- at least in the massless limit, see
 \cite{Jegerlehner:2005fs,Pasechnik:2005ae,Kataev:2013vua}.}

To establish a relation of the above spectral density to the pion-photon
TFF, we propose to saturate the three-point correlation function by means
of resonances, notably, the pion state plus a continuum of all higher
resonances.
Then, assuming the validity of the global quark-hadron duality,
we can express the contribution of all higher resonances as that part
of the total integral in the ASR, i.e., the integral
over the same spectral density $A_{3}(s,Q^{2};m^{2})$,
which starts from some lower limit $s_0$.
Note that, in general, $s_0$ can depend on $Q^2$ and has the
meaning of the pion duality interval.
Taking into account that the pion resonances contribute via the
pion-photon TFF, i.e., Eq.\ (\ref{eq:matrix-element}), and recalling
that the original ASR is given in terms of Eq.\ (\ref{eq:asr}), we
finally obtain the following SR
\begin{equation}
  \pi f_{\pi}F_{\pi\gamma}(Q^2)
=
  \frac{1}{2\pi}N_c C-\int_{s_0}^{\infty} A_{3}(s,Q^{2};m_i^{2}) ds \, .
 \label{eq:qhd3}
\end{equation}
Here, the pion decay constant $f_\pi$ is defined as the matrix element
$\langle 0|J_{\alpha 5}(0) |M(p)\rangle=i p_\alpha f_\pi $
and the pion-photon TFF $F_{\pi\gamma}(Q^2)$
is given by Eq.\ (\ref{eq:matrix-element}).

The contribution to the spectral density $ A_{3}(s,Q^{2};m^{2})$ for a
given flavor $q$ is \cite{Horejsi:1994aj},
\begin{equation}
  A_3^{(q)}(s,Q^{2};m_q^{2})
=
  \frac{e_q^2}{2\pi}
  \frac{1}{(Q^2+s)^2}
  \left(Q^2R+2m_q^2 \ln \frac{1+R}{1-R}\right)\, ,
\label{eq:a3}
\end{equation}
where
$R(s,m_q^2)=\sqrt{1-\frac{4m_q^2}{s}}$
$m_q$ is the quark mass for flavor $q$.
Neglecting the quark mass, we obtain from (\ref{eq:qhd3}) and
(\ref{eq:a3}) the following expression for the pion-photon TFF,
\begin{align}
  F_{\pi\gamma}(Q^2)
=
  \frac{1}{2\sqrt{2}\pi^2f_{\pi}}
  \frac{s_0(Q^2)}{s_0(Q^2)+Q^2} \, ,
\label{eq:f3m}
\end{align}
where $f_{\pi}=0.134$~MeV is the pion decay constant.

Note that the ASR relation is valid for any value of $Q^2$
(given that the total integral does not depend on it).
But, in addition, also relation (\ref{eq:f3m}) is valid for all $Q^2$
values, hence defining an exact property of the TFF.
On the most basic level, one obviously recovers for $Q^2=0$ the
chiral limit $\frac{1}{2\sqrt{2}\pi^2f_{\pi}}F_{\pi\gamma}(Q^2=0)=1$.
Moreover, the continuation of (\ref{eq:f3m}) from the
spacelike region to the timelike domain can also be performed
\cite{Klopot:2013laa}.
This procedure does not violate the exactness of the ASR, viz.,
Eq.\ (\ref{eq:asr}), provided one assumes that $s_0$ has the most
general form which allows for a dependence on the momentum transfer
\cite{Melikhov:2012qp}.
Let us assume that, though $s_0$ resembles the so-called continuum
threshold parameter in QCD sum rules, it may vary with $Q^2$ ---
like in the KOT approach.
In addition, it may also have different values in the two- and
three-point correlators.
Be that as it may, as long as we do not fix the form of $s_0(Q^2)$,
the  ASR result for the pion-photon TFF given by Eq.\ (\ref{eq:f3m})
remains an exact relation.
One has to fix the form of $s_0(Q^2)$ only if one is interested in the
numerical value of this quantity.
This may eventually entail some inaccuracy, like in the KOT approach.

On the pretext that $s_0$ is a continuum threshold parameter,
we may suppose that in first approximation $s_0$ is a constant.
With this assumption,
$s_0$ can be determined from Eq.\ (\ref{eq:f3m}) in the limit
$Q^2\to \infty$ of the spacelike ASR (see \cite{Klopot:2010ke}).
As in the asymptotic limit, the validity of the factorization theorem
should be completely restored, so that one easily obtains
$s_0=4\pi^2f_\pi^2\simeq 0.67$ GeV$^2$.
This expression coincides with the one found earlier from
a two-point correlator analysis in \cite{Radyushkin:1995pj} and
is also close to the numerical value estimated using
two-point sum rules \cite{Shifman:1978by}.
In this way, we find that the Brodsky-Lepage interpolation formula
\cite{Brodsky:1981rp}
(which corresponds to the one-loop approximation of the ASR)
holds in the timelike region as well.\footnote{The similarity between
the Brodsky-Lepage interpolation formula in the spacelike region and
the vector-dominance model in the timelike region is well known,
see, e.g., \cite{CELLO91}.}
It is also worth mentioning that an expression similar to 
Eq.\ (\ref{eq:f3m}) was derived in \cite{Masjuan:2012sk}
by employing an extension of VMD ideas to include higher resonances
in combination with the correct large-$Q^2$ behavior of the TFF.

In this paper we want to scrutinize the assumption that $s_0$ can be
supposed to be close to a constant.
To this end, we will assume that $s_0$ is some arbitrary (but smooth)
function of $Q^2$ and then compare the ASR relation for the TFF,
cf.\ (\ref{eq:f3m}), with the predictions obtained within the BMPS
LCSR-based analysis in
\cite{Stefanis:2012yw,Mikhailov:2010ud,Bakulev:2011rp}.

\section{Interplay between lightcone and anomaly sum rules}
\label{sec:lcsr-asr}
\begin{figure}[t]
\centering
\includegraphics[width=0.85\textwidth]{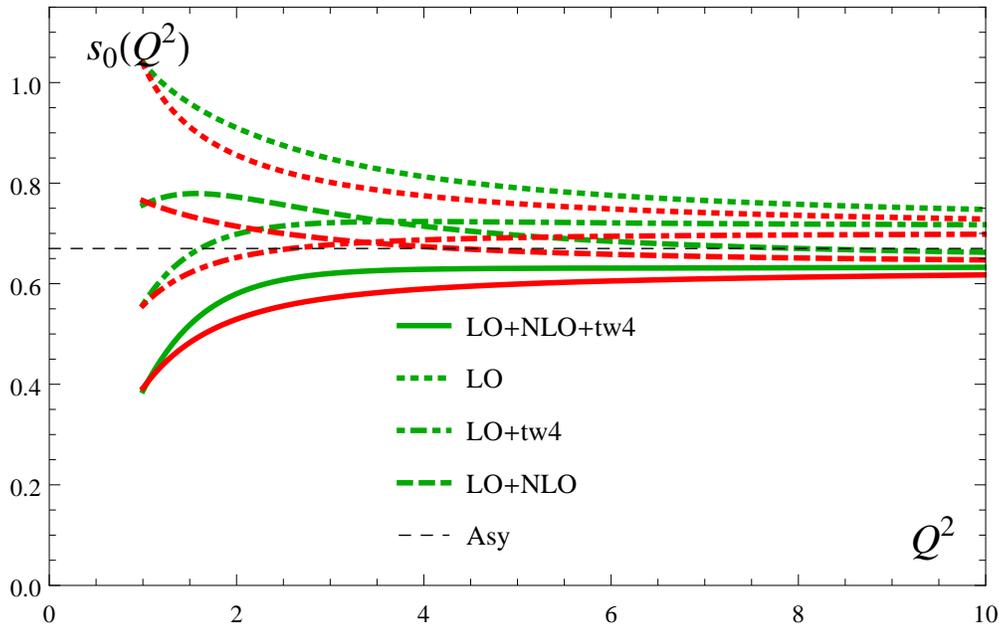}
\caption{(Color online).
Anomaly SR threshold parameter $s_0(Q^2)$ extracted from the
comparison of the pion-photon TFF obtained from the anomaly SR
and from LCSRs at four levels of accuracy:
(i) dotted lines --- LO result;
(ii) dashed lines --- LO+NLO;
(iii) dashed-dotted lines --- LO with twist four contribution included;
(iv) thick solid line --- LO+NLO+twist-4.
The horizontal dashed line represents the threshold value
$s_0=4\pi^2f_\pi^2\simeq 0.67$ GeV$^2$.
The green lines show the results for the BMS pion DA model \cite{BMS01},
whereas the red lines display the analogous results for the
shorttailed platykurtic pion DA \cite{Stefanis:2014yha}.
Note that from top to bottom, all red curves appear always
below the corresponding green ones.}
\label{fig:s0}
\end{figure}

\begin{figure}[t]
\centering
\includegraphics[width=0.75\textwidth]{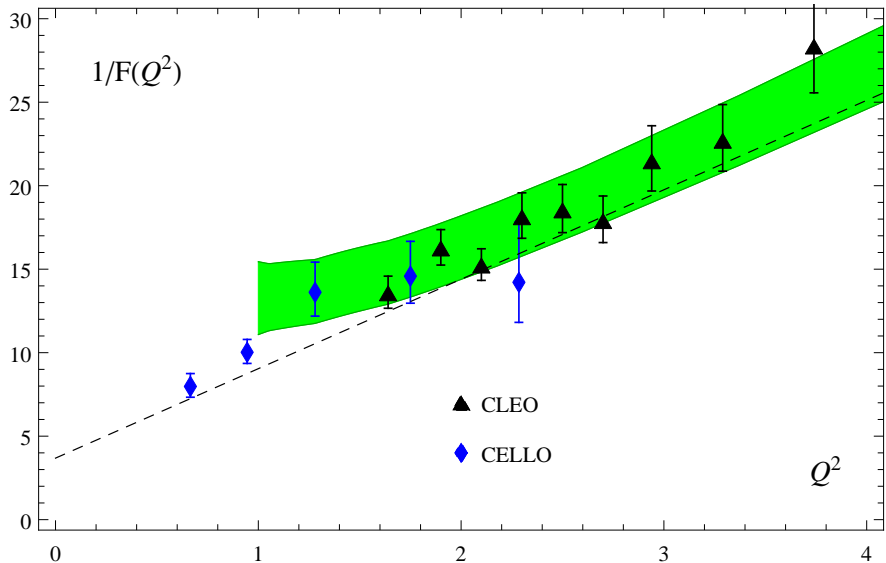}
\caption{(Color online).
Inverse pion-photon TFF obtained from two
different sum-rule approaches: LCSR (broad green band)
and linear behavior from anomaly SR (dashed line).
The experimental data are taken from CELLO
\cite{CELLO91} --- blue diamonds and
CLEO \cite{CLEO98} --- black triangles.}
\label{fig:inverseFF}
\end{figure}

In this section, we will work out the connection between the two
sum-rule approaches considered above.
Our comparative analysis will be carried out in three steps:

\begin{enumerate}
\item
Presuming that the threshold $s_0(Q^2)$ is some definite (but unknown)
function of the momentum, we estimate its value by comparing
(\ref{eq:f3m}) with (\ref{eq:LCSR-FQq})
within the validity range of the LCSR approach
($Q^2\gtrsim 1$~GeV$^2$).
The outcome of this comparison is shown in Fig.\ \ref{fig:s0}.
The various lines correspond to different contributions, taken into
account step by step by means of LCSRs.
Their meaning is explained from top to bottom.
Line one (dotted) shows the LO result, while the next lower one (dashed)
illustrates an analogous result when also the NLO term has been included
in the TFF.
The third line (dashed-dotted) represents the prediction which includes
only the LO term and the twist-four contribution, with the NLO term being
excluded.
Finally, the lowest curve (thick solid) displays the total result which
comprises the NLO radiative contribution as well as the twist-four term.
In addition, also the limiting case $s_0=0.67$~GeV$^2$, obtained from ASRs
under the assumption that $s_0$ is constant, is shown in the form of a
horizontal dashed line.
The analysis was carried out using two different models for the pion DA:
the bimodal BMS DA \cite{BMS01} (green lines)
and the shorttailed platykurtic DA \cite{Stefanis:2014yha} (red lines
--- always below the green ones).

One observes from this figure that below 2-3~GeV$^2$ the various curves
start to deviate significantly from each other calling for more
detailed considerations to be addressed in the next item.
On the other hand, it is worth noting that above
$Q^2 \gtrsim 3$~GeV$^2$, $s_0(Q^2)$ starts to scale with $Q^2$
and has a constant magnitude that depends on the actual
approximation of the LCSR.
Let us note that the total result for both, the BMS pion DA
and the platykurtic DA, leads to a $s_0$ value rather far from the constant
$0.67$~GeV$^2$ used in the KOT approach --- even in the intermediate region
$Q^2<10$~GeV$^2$.
On the other hand, in the NLO approximation of the LCSR
(LO+NLO without twist 4 contribution), expressions (\ref{eq:f3m}) and
(\ref{eq:LCSR-FQq}) numerically coincide, thus confirming the
assumption that $s_0$ is close to a constant.
The explanation of this coincidence will be discussed
shortly.

\item
Let us concentrate on the low-momentum region below 3~GeV$^2$.
Here, the role of higher-order corrections to the LCSR,
both perturbative and nonperturbative, becomes significant,
whereas the ASR result (according to the KOT assumption) does not
change, being $Q^2$ independent.
This becomes evident from the comparison of the LO TFF
prediction with the one which includes the NLO correction, employing
in both cases the BMS DA within the BMPS framework.
As one can see from Fig.\ \ref{fig:s0}, the LO approximation to
the LCSR leads to an unreliable dependence on $Q^2$, even if the
twist-four contribution is incorporated.
It is only when the NLO term is taken into account that the result
stabilizes, leading to a $s_0$ close to a constant.
This behavior reflects the fact that at such virtualities perturbative,
i.e., radiative, corrections to the LCSR are vital.
It is remarkable that using instead the shorttailed platykurtic DA,
one finds that $s_0$ almost coincides with the result derived in
the KOT ASR in the entire range of $Q^2$.
In contrast, incorporating only the twist-four contribution renders
also this result worse, remaining bad even after the inclusion of the NLO
term, as already mentioned.
In both cases, the value of $s_0$ decreases appreciably and, as a result,
stability against the $Q^2$ variation gets worse as well.

One should note, however, that effects stemming from large-distance
dynamics are accounted for by the anomaly relation itself ``as a whole'',
implying that to compare its prediction with that obtained with LCSRs, one
should include in the latter the sum over the whole infinite series
of the nonperturbative corrections \cite{Teryaev:2013qba}.
The absence of a nonperturbative correction to the ASR according to
t'Hooft's principle may indicate that a strong cancellation of
nonperturbative corrections takes place.
Indeed, estimations in \cite{Bakulev:2011rp} show that the next term in the
nonperturbative expansion (the twist-six contribution) has the
opposite sign relative to the twist four term and should therefore grossly
cancel it's contribution.
Moreover, one may provide additional arguments that a significant
cancellation of the nonperturbative corrections may indeed take place.
In fact, if one expands the exact ASR result (\ref{eq:f3m}) for the
pion TFF in a series in terms of the ratio $s_0/Q^2$, one obtains an
expression which could be regarded as an infinite-series expansion of the
nonperturbative corrections.
One can easily see that, for the physically reliable case of $s_0$ being
close to a constant (or at least a not strongly oscillating function
of $Q^2$), the series should be an alternating one.

This clearly means that at momentum values around
$Q^2=1-2$~GeV$^2$, a strong cancelation of the perturbative corrections
should take place.
Thus, we may come to the conclusion that the ASR result should be
compared with the prediction derived from the LCSR only when the
NLO term is included while the twist-four contribution is excluded.
For this case one may further conclude from Fig.\ \ref{fig:s0}
that the comparison with the BMPS result (using the BMS DA) supports
the assumption that the pion duality interval $s_0(Q^2)$ in the KOT
approach is close to a constant (second line from the top in this
figure), implying that the approximation $s_0=0.67$~GeV$^2$ is
reasonable  within a 10 percent accuracy.
Note that employing the shorttailed platykurtic pion DA,
this agreement becomes even better.

\item
For a more accurate analysis of the TFF, it is convenient
to analyze the behavior of the pion-photon TFF at
$q^2=-Q^2 \rightarrow 0$
in terms of the ratio
$R\equiv F_{\pi\gamma}(q^2)/F_{\pi\gamma}(0)$.
The dimensionless slope and curvature parameters at $q^2=0$,
are defined by
$a=m_{\pi}^2\partial R/\partial q^2|_{q^2=0}$
and
$b=\frac{1}{2}m_{\pi}^4\partial^2R/\partial (q^2)^2|_{q^2=0}$,
respectively, and have for the pion the following values:
$a_{\pi}= m_{\pi}^2/s_0 = 0.027$ and
$b_{\pi}=m_{\pi}^4/s_0^2=0.73\times 10^{-3}$.
The slope $a_\pi$ was recently determined in \cite{Masjuan:2012wy} 
by means of Pad\'e approximants as fitting functions in the analysis 
of the spacelike experimental data of the pion-photon TFF and was 
found to be $a_\pi\simeq 0.0324$.
A more recent dispersive analysis in \cite{Hoferichter:2014vra} extracted
at vanishing momentum transfer the value $a_\pi\simeq 0.0307$.
Both results for the slope are (within the range of the estimated errors
omitted here) compatible with each other and also in good agreement with
the ASR result within the $10\%$ margin of accuracy at the lower limit
discussed before.
From (\ref{eq:f3m}) one can easily see that the slope of the TFF
at $q^2=0$ depends only on the value $s_0$ due to $dR/dq^2=1/s_0$.
Thus, we can conclude that, though $s_0(q^2)$ can be treated as constant
to the level of 10 percent precision, a more accurate treatment would be
to start at the value $s_0=0.61$~GeV$^2$ for $q^2=0$ and approach
the asymptotic value $s_0=0.67$~GeV$^2$ at the scale $-q^2=Q^2=1-2$~GeV$^2$.
It is worth noting that the value $s_0=0.61$~GeV$^2$ at $q^2=0$ coincides
with the $\rho$-meson mass squared.
The slope $a_{\pi}$ can be related to the pion radius by the trivial relation
$
 (\langle r_{\pi}\rangle)^2
=
 6a_{\pi}/m_{\pi}^2
$
(see \cite{Bernstein:2015sac} for a discussion).
In this way we find the pion radius for
$s_0=0.61$~GeV$^2$ to be $0.62$~fm, while for the asymptotic value
$s_0=0.67$~GeV$^2$ the result is slightly smaller and about $0.59$~fm.
Both values are in accord with the pion radius
derived from considerations based on the vector-meson dominance,
whereas the pion radius extracted from the DS-based approach 
\cite{Chang:2013pq} turns out to be somewhat larger, $0.68$~fm.

As regards the curvature parameter
$b_\pi$ at $q^2=0$,
it was computed in the works mentioned above and found to be
$b_\pi\approx 0.0011$ --- see \cite{Masjuan:2012wy,Hoferichter:2014vra}
for the corresponding computational details.
Comparing this value with the ASR result shows that agreement can be
achieved at
$s_0=0.61$~GeV$^2$ (with $q^2=0$), allowing for the $s_0$ derivative
to be negative, $ds_0/dq^2=-ds_0/dQ^2= -0.25$.
In summary, it is reasonable to expect that $s_0$ is $0.61$~GeV$^2$ at
the zero point and grows with $Q^2$ up to the asymptotic
value $s_0=0.67$~GeV$^2$ at scales on the order of $1-2$~GeV$^2$.

\item
It was noted in Ref.\ \cite{Klopot:2013laa} that analytic continuation
of the ASR approach to the timelike region leads to a pole in the TFF
at $q^2=s_0$, which is numerically close to the $\rho$ meson mass squared,
i.e., $m_\rho^2\simeq 0.59$~GeV$^2$.
This pole can be traced back to the analytic continuation of the ASR into
the timelike domain and arises because in the spacelike region the TFF is
proportional to $s_0/(s_0-q^2)$.
As a result, the inverse TFF $1/F(Q^2)$ should show a quasi linear behavior
at moderately large $Q^2$.
Furthermore, it should also agree with the well-known limit of the TFF for
two real photons (i.e, the two-photon pion decay width).
Because the LCSRs are not reliably applicable at very low momenta
below, say, below 1~GeV$^2$, it looks promising to compare the KOT
results at low $Q^2=-q^2>0$ in the spacelike region with the experimental
data and to consider a possible (linear) interpolation of the BMS prediction
to the well-known limit of the TFF for two real photons
(i.e., the two-photon decay width, which clearly does not depend on the $s_0$ value).
Therefore, we show in Fig.\ \ref{fig:inverseFF} the predictions for the
inverse TFF $1/F(Q^2)$
obtained with the KOT approach with a constant $s_0$ value
(solid line) and also with the BMPS method, contrasting them with the
experimental data at low $Q^2$.
The BMPS predictions are shown by means of a shaded (green)
area which indicates its accuracy range.
The slight excess of the BMS results can be interpreted as a
reflection of the small ($\sim 10\%$) variation of $s_0(Q^2)$ at small $Q^2$
and uncertainties originating from the LCSRs themselves at small $Q^2$ values.
\end{enumerate}

We see from Fig.\ \ref{fig:inverseFF} that the ASR result
(solid line) is in good agreement with the experimental data and exactly
corresponds to the two-photon decay width at $Q^2=0$.
On the other hand, the LCSR prediction also conforms with experiment ---
starting from about 2~GeV$^2$ --- and allows for a linear
interpolation of $1/F(Q^2)$ to the two-photon decay width at $Q^2=0$.
However, the LCSR result for the inverse TFF $1/F(Q^2)$ starts
deviating from the linear interpolation formula at lower momenta
in the region 1-1.5~GeV$^2$ where the LCSRs are supposed to be still
reliable.
Indeed, a smooth continuous interpolation (including the derivative) to
the limit of the two-photon decay width is not showing a linear behavior.
This indicates that $s_0$ may deviate from a constant at small
$Q^2<1.5$~GeV$^2$ at the level of $10-20 \% $.
This deviation may be the result of uncertainties of the LCSR method
at small momentum scales currently under investigation \cite{MPS15},
see also \cite{Bakulev:2011rp}, where the interplay between the main
NNLO contribution ($\beta_0$-part) and the twist-six term is discussed.
If this deviation would be confirmed by further analysis,
it would be very interesting to estimate its analytic form,
especially bearing in mind that, after the analytic continuation to
the timelike region (see Ref.\ \cite{Klopot:2013laa}), the deviation
from linearity could be related to the value of the widths of vector mesons.
Unfortunately the current knowledge of the intrinsic uncertainties
of the LCSR predictions in the low $Q^2$ domain is quite limited to allow
a precise estimation of this deviation.
We hope that the dedicated low $Q^2$ analysis of the LCSRs in \cite{MPS15}
will provide more clues.

\section{Conclusions}
\label{sec:concl}
In this work we performed a comparative analysis of the pion-photon TFF
computed with two different types of sum rules: LCSRs and ASRs.
The first method interpolates correctly the behavior of the
electromagnetic pion-photon transition form factor for a highly virtual
and a real photon from the ultraviolet limit of QCD down to typical
hadronic scales of about $1-2$~GeV, close to the dipole formula.
The second method is based on the chiral anomaly and is exact even when
both photons have vanishing virtualities.
The objective was to match the key parameters of these approaches in
the low-$Q^2$ domain thus, in some sense, ``unifying'' their predictions.
To this end, we used within the BMPS LCSR approach two different types
of pion DAs.
The classic bimodal BMS DAs \cite{BMS01} and the unimodal shorttailed
platykurtic DA \cite{Stefanis:2014nla,Stefanis:2015qha}.
The common key element of both DAs is the strong suppression of the
kinematic endpoint regions $x=0,1$.
This behavior distinguishes them from many other model DAs, ranging
from the bimodal endpoint-concentrated Chernyak-Zhitnitsky
\cite{Chernyak:1983ej} DA to the recently proposed DSE-based DAs
which have broad unimodal profiles with strongly enhanced tails.
The main results of our analysis have been presented graphically in
Fig.\ \ref{fig:s0} and Fig.\ \ref{fig:inverseFF}.
From the first figure we observe that the pion-photon TFF, computed
with the platykurtic DA within the LCSR framework, strongly resembles
the analogous result computed with the ASRs.
However, this mutual consistency is best only if the twist-four
contribution in the former calculation is excluded, arguing that the
ASR prediction inherently embodies not only the first
next-to-leading-order twist term, but a whole series of such
contributions.
Moreover, from this agreement one can infer that the threshold
parameter $s_0$ in the ASR approach is a constant with a matching value
$s_0=0.67$~GeV$^2$ at the accuracy level of $10\%$.
On the other hand, inspection of Fig.\ \ref{fig:inverseFF} reveals that
strong cancelations of the radiative corrections are needed in order to
produce the linear behavior of the displayed ASR result (dashed line).
But, from the other side, if the observed deviation is real it might
indicate a dependence of $s_0$ on the large photon virtuality $Q^2$.
From our combined analysis we found that the value $s_0$ can vary
between the value $0.61$~GeV$^2$ at $Q^2=0$ and the asymptotic value
$0.67$~GeV$^2$ at the scale $1-2$~GeV$^2$.
We also found by employing the ASR approach that the pion radius can
be estimated to be in the interval $0.59-0.62$~fm.

\section*{Acknowledgments}
We thank Y.~N.~Klopot and S.~V.~Mikhailov for useful discussions
and comments.
This work was partially supported by the Heisenberg--Landau Program
(Grant 2015),
the Russian Foundation for Basic Research under Grants No.\ 14-01-00647
and No.\ 15-52-04023, the JINR-BelRFFR grant F14D-007,
the Major State Basic Research Development Program in China
(No.\ 2015CB856903),
and the National Natural Science Foundation of China
(Grant Nos.\ 11575254 and 11175215).


\begin{thebibliography}{10}
\bibitem{Aubert:2009mc}
B. Aubert {\it et~al.}, Phys. Rev.
\textbf{D80},  052002  (2009). 

\bibitem{Lepage:1980fj}
G.~P. Lepage and S.~J. Brodsky, Phys. Rev.
\textbf{D22},  2157  (1980). 

\bibitem{Brodsky:1981rp}
S.~J. Brodsky and G.~P. Lepage, Phys. Rev.
\textbf{D24},  1808  (1981). 

\bibitem{Uehara:2012ag}
S. Uehara {\it et~al.}, Phys. Rev.
\textbf{D86},  092007  (2012). 

\bibitem{MS09}
S.~V. Mikhailov and N.~G. Stefanis, Nucl. Phys.
\textbf{B821},  291  (2009). 

\bibitem{Radyushkin:2009zg}
A.~V. Radyushkin, Phys. Rev.
\textbf{D80},  094009  (2009). 

\bibitem{Polyakov:2009je}
M.~V. Polyakov, JETP Lett.
\textbf{90},  228  (2009). 

\bibitem{Stefanis:2015qha}
N.~G. Stefanis and A.~V. Pimikov, Nucl. Phys.
\textbf{A945},  248  (2016). 

\bibitem{ER80}
A.~V. Efremov and A.~V. Radyushkin, Phys. Lett.
\textbf{B94},  245  (1980). 

\bibitem{BBK89}
I.~I. Balitsky, V.~M. Braun, and A.~V. Kolesnichenko, Nucl. Phys.
\textbf{B312},  509  (1989). 

\bibitem{Kho99}
A. Khodjamirian, Eur. Phys. J.
\textbf{C6},  477  (1999). 

\bibitem{BMS02}
A.~P. Bakulev, S.~V. Mikhailov, and N.~G. Stefanis, Phys. Rev.
\textbf{D67},  074012  (2003). 

\bibitem{Bakulev:2003cs}
A.~P. Bakulev, S.~V. Mikhailov, and N.~G. Stefanis, Phys. Lett.
\textbf{B578},  91  (2004). 

\bibitem{BMS05lat}
A.~P. Bakulev, S.~V. Mikhailov, and N.~G. Stefanis, Phys. Rev.
\textbf{D73},  056002  (2006). 

\bibitem{Mikhailov:2010ud}
S.~V. Mikhailov, A.~V. Pimikov, and N.~G. Stefanis, Phys. Rev.
\textbf{D82},  054020  (2010). 

\bibitem{Bakulev:2011rp}
A.~V. Bakulev, S.~V. Mikhailov, A.~V. Pimikov, and N.~G. Stefanis, Phys. Rev.
\textbf{D84},  034014  (2011). 

\bibitem{BMPS12}
A.~P. Bakulev, S.~V. Mikhailov, A.~V. Pimikov, and N.~G. Stefanis, Phys. Rev.
\textbf{D86},  031501  (2012). 

\bibitem{Stefanis:2012yw}
N.~G. Stefanis, A.~V. Bakulev, S.~V. Mikhailov, and A.~V. Pimikov, Phys. Rev.
\textbf{D87},  094025  (2013). 

\bibitem{Khodjamirian:2009ib}
A. Khodjamirian, Int. J. Mod. Phys.
\textbf{A25},  513  (2010). 

\bibitem{ABOP10}
S.~S. Agaev, V.~M. Braun, N. Offen, and F.~A. Porkert, Phys. Rev.
\textbf{D83},  054020  (2011). 

\bibitem{ABOP12}
S.~S. Agaev, V.~M. Braun, N. Offen, and F.~A. Porkert, Phys. Rev.
\textbf{D86},  077504  (2012). 

\bibitem{Klopot:2010ke}
Y.~N. Klopot, A.~G. Oganesian, and O.~V. Teryaev, Phys. Lett.
\textbf{B695},  130  (2011). 

\bibitem{Klopot:2011qq}
Y.~N. Klopot, A.~G. Oganesian, and O.~V. Teryaev, Phys. Rev.
\textbf{D84},  051901  (2011). 

\bibitem{Klopot:2011ai}
Y.~N. Klopot, A.~G. Oganesian, and O.~V. Teryaev, JETP Lett.
\textbf{94},  729  (2011). 

\bibitem{Klopot:2012hd}
Y.~N. Klopot, A.~G. Oganesian, and O.~V. Teryaev, Phys. Rev.
\textbf{D87},  036013  (2013). 

\bibitem{Dolgov:1971ri}
A.~D. Dolgov and V.~I. Zakharov, Nucl. Phys.
\textbf{B27},  525  (1971). 

\bibitem{Horejsi:1985qu}
J. Horejsi, Phys. Rev.
\textbf{D32},  1029  (1985). 

\bibitem{Horejsi:1994aj}
J. Horejsi and O.~V. Teryaev, Z. Phys.
\textbf{C65},  691  (1995). 

\bibitem{Klopot:2013laa}
Y.~N. Klopot, A.~G. Oganesian, and O.~V. Teryaev, JETP Lett.
\textbf{99},  679  (2014). 

\bibitem{BMS01}
A.~P. Bakulev, S.~V. Mikhailov, and N.~G. Stefanis, Phys. Lett.
\textbf{B508},  279  (2001);
\textit{ibid.} \textbf{B590} (2004), 309 (Erratum).

\bibitem{SY99}
A. Schmedding and O. Yakovlev, Phys. Rev.
\textbf{D62},  116002  (2000). 

\bibitem{MR86}
S.~V. Mikhailov and A.~V. Radyushkin, JETP Lett.
\textbf{43},  712  (1986). 

\bibitem{MR86ev}
S.~V. Mikhailov and A.~V. Radyushkin, Nucl. Phys.
\textbf{B273},  297  (1986). 

\bibitem{MR89}
S.~V. Mikhailov and A.~V. Radyushkin, Sov. J. Nucl. Phys.
\textbf{49},  494  (1989). 

\bibitem{MR90}
S.~V. Mikhailov and A.~V. Radyushkin, Sov. J. Nucl. Phys.
\textbf{52},  697  (1990). 

\bibitem{MS93}
S.~V. Mikhailov, Phys. Atom. Nucl.
\textbf{56},  650  (1993). 

\bibitem{BR91}
A.~P. Bakulev and A.~V. Radyushkin, Phys. Lett.
\textbf{B271},  223  (1991). 

\bibitem{MR92}
S.~V. Mikhailov and A.~V. Radyushkin, Phys. Rev.
\textbf{D45},  1754  (1992). 

\bibitem{Melic:2002ij}
B. Meli\'c, D. M\"uller, and K. Passek-Kumeri\v{c}ki, Phys. Rev.
\textbf{D68},  014013  (2003). 

\bibitem{Stefanis:2014nla}
N.~G. Stefanis, Phys. Lett.
\textbf{B738},  483  (2014). 

\bibitem{Stefanis:2014yha}
N.~G. Stefanis, S.~V. Mikhailov, and A.~V. Pimikov, Few Body Syst.
\textbf{56},  295  (2015). 

\bibitem{Chang:2013pq}
L. Chang {\it et~al.}, Phys. Rev. Lett.
\textbf{110},  132001  (2013). 

\bibitem{Gao:2014bca}
F. Gao {\it et~al.}, Phys. Rev.
\textbf{D90},  014011  (2014). 

\bibitem{SSK99}
N.~G. Stefanis, W. Schroers, and H.-C. Kim, Phys. Lett.
\textbf{B449},  299  (1999). 

\bibitem{SSK00}
N.~G. Stefanis, W. Schroers, and H.-C. Kim, Eur. Phys. J.
\textbf{C18},  137  (2000). 

\bibitem{Adler:1969gk}
S.~L. Adler, Phys. Rev.
\textbf{177},  2426  (1969). 

\bibitem{'tHooft:1980xb}
G. 't~Hooft {\it et~al.}, NATO Sci.Ser.B
\textbf{59},  pp.1  (1980). 

\bibitem{Jegerlehner:2005fs}
F. Jegerlehner and O.~V. Tarasov, Phys. Lett.
\textbf{B639},  299  (2006). 

\bibitem{Pasechnik:2005ae}
R.~S. Pasechnik and O.~V. Teryaev, Phys. Rev.
\textbf{D73},  034017  (2006). 

\bibitem{Kataev:2013vua}
A.~L.~Kataev, JHEP 
\textbf{1402},  092  (2014). 

\bibitem{Melikhov:2012qp}
D. Melikhov and B. Stech, Phys. Lett.
\textbf{B718},  488  (2012). 

\bibitem{Radyushkin:1995pj}
A.~V. Radyushkin, Acta Phys. Polon.
\textbf{B26},  2067  (1995). 

\bibitem{Shifman:1978by}
M.~A. Shifman, A. Vainshtein, and V.~I. Zakharov, Nucl. Phys.
\textbf{B147},  448  (1979). 

\bibitem{CELLO91}
H.~J. Behrend {\it et~al.}, Z. Phys.
\textbf{C49},  401  (1991). 

\bibitem{Masjuan:2012sk}
P.~Masjuan, E.~Ruiz Arriola, and W.~Broniowski, Phys. Rev.
\textbf{D87},  014005  (2013). 

\bibitem{CLEO98}
J. Gronberg {\it et~al.}, Phys. Rev.
\textbf{D57},  33  (1998). 

\bibitem{Teryaev:2013qba}
O.~V. Teryaev, Nucl. Phys. Proc. Suppl.
\textbf{245},  195  (2013). 

\bibitem{Masjuan:2012wy}
P. Masjuan, Phys. Rev.
\textbf{D86},  094021 (2012). 

\bibitem{Hoferichter:2014vra}
M. Hoferichter {\it et~al.}, Eur. Phys. J.
\textbf{C74},  3180  (2014). 

\bibitem{Bernstein:2015sac}
A.~M. Bernstein,  in {\em {8th International Workshop on Chiral Dynamics (CD
2015) Pisa, Italy, June 29-July 3, 2015}}
(PUBLISHER, ADDRESS, 2015). 

\bibitem{MPS15}
S.~V. Mikhailov, A.~V. Pimikov, and N.~G. Stefanis,
\uppercase{B}ochum preprint
RUB-TPII-04/2015, work in progress. 

\bibitem{Chernyak:1983ej}
V.~L. Chernyak and A.~R. Zhitnitsky, Phys. Rept.
\textbf{112},  173  (1984). 

\end{thebibliography}

\end{document}